\documentclass{nic-series}

\begin{document} 

\title{Effective hamiltonian approach for strongly correlated lattice models}

\author{Sandro Sorella}

\institute{ Istituto Nazionale per la Fisica della Materia, I-34014  Trieste Italy\\ 
  SISSA, Via Beirut n.2-4, I-34014 Trieste, Italy\\\email{sorella@sissa.it}
          }
\maketitle

\begin{abstracts}
We review  a recent  approach 
for the simulation of many-body interacting systems based on an  efficient 
generalization of the Lanczos method  for Quantum Monte Carlo simulations.
This technique allows to perform systematic corrections to a given variational 
wavefunction, that allow to estimate  exact energies and correlation 
functions, whenever the starting variational wavefunction is a qualitatively correct 
description of the ground state.
The stability of the variational wavefunction against possible phases 
, not described  at the variational level can be tested  by using the  
''effective Hamiltonian'' approach. In fact  Monte Carlo methods,
 such as the ''fixed node approximation''
 and the present ''generalized Lanczos technique'' (Phys. Rev. B 64,024512, 2001) allow 
to obtain exact ground state properties  of an effective Hamiltonian, chosen to be as close 
as possible to the exact Hamiltonian, thus yielding  the most reasonable estimates
of correlation functions.  We also describe a simplified one-parameter scheme 
that  improve substantially the 
efficiency of the generalized Lanczos method. 
This is  tested on the $t-J$ model,  with  a special effort   
to obtain accurate pairing correlations, and provide  a  possible 
non-phonon mechanism for High temperature superconductivity. 
\end{abstracts}

\section{Introduction}
\label{sec_motive}

Despite the tremendous progress of computer performances 
the general task of determining the ground state 
wavefunction of a many-electron system   is  still  far from being settled 
For instance, 
even for simplified models on a lattice, there is no general consensus 
on the ground state properties of a system  of about $~100$ electrons on
$L~100$ sites.
The most striking example is the so called $t-J$ model:
This model  is still a subject 
of intense numerical studies, due to its possible relevance 
for  High Tc superconductivity\cite{anderson,zhang}. The Hamiltonian reads:
\begin{equation}\label{tj}
\hat{H}= J \sum_{\langle i,j \rangle} \left ( \hat{{\bf S}}_i \cdot \hat{{\bf S}}_j -
\frac{1}{4} \hat{n}_i \hat{n}_j \right )
-t \sum_{\langle i,j \rangle, \sigma}
{\tilde c}^{\dag}_{i,\sigma} {\tilde c}_{j,\sigma},
\end{equation}
where ${\tilde c}^{\dag}_{i,\sigma}=\hat{c}^{\dag}_{i,\sigma}
\left ( 1- \hat{n}_{i,\bar \sigma} \right )$,
$\hat{n}_i= \sum_{\sigma} \hat{n}_{i,\sigma}$ is the electron
density on site $i$,
$\hat{{\bf S}}_i=\sum_{\sigma,\sigma^{\prime}}
{\tilde c}^{\dag}_{i,\sigma} {\bf \tau}_{\sigma,\sigma^{\prime}}
{\tilde c}_{i,\sigma^{\prime}}$ is the spin operator and
${\bf \tau}_{\sigma,\sigma^{\prime}}$ are Pauli matrices.
In the following we consider  $N$ electrons on $L$ sites, with  
 periodic boundary conditions,(PBC), in order  to minimize size effects. 

After many years of intense numerical and theoretical efforts  
there is no general consensus on the properties  
of this simple Hamiltonian and of the related Hubbard model.
In particular according to density matrix renormalization group (DMRG) studies \cite{white}, d-wave superconductivity is 
not stable in this model, whereas a ground state  non uniform  in density (with so called  ''stripes'') 
is found.
Several QMC studies provide   controversial results, most of them indicating  a  
superconducting behavior, and some of them\cite{tklee}, indicating the opposite.

The reason of the above controversy, can be easily explained within the straightforward 
variational approach. Whenever a model Hamiltonian cannot be solved exactly
 either numerically (with no sign 
problem) or analytically,  the  most general and reasonable approach  
is  an approximate minimization of the energy within a particular class of wavefunctions, for instance 
also DMRG can be considered a variational approach with a particularly complicated variational 
wavefunction obtained by DMRG iterations.
However,  within the variational approach,  one faces the following problem:
for large system size $L$ the gap to the first excited state scales  generally to zero quite rapidly with 
$L$.  Thus   between the ground state energy and the variational energy there maybe a very large number 
of states with completely different correlation functions.
In this way one can generally obtain different variational wavefunctions with almost similar 
energy per site, but with completely different correlation functions.
It is  easily understood that,  within a straightforward variational technique,
  there 
is no hope to obtain sensible results for large system size,
 unless for for system with a  finite gap to all 
excitations, such as spin liquid\cite{rvb},  or band insulators..

In the following we are trying to argue that a possible solution 
to the previous limitation of the variational technique  is 
provided by what we call in 
the following ''the effective Hamiltonian approach''.

This approach relies on  the following assumption:

''Among similar Hamiltonians with local interactions the ground state correlation functions 
  depend weakly on the details of the Hamiltonian, in a sense that similar Hamiltonians 
should provide similar correlation functions'''.
In this way  the ground state of an effective  Hamiltonian (such as the fixed node 
Hamiltonian\cite{ceperley}) that 
can be solved exactly by Quantum Monte Carlo schemes can be used as a variational state of the 
desired Hamiltonian, in this way providing not only a good variational energy but the most 
reasonable estimate of correlation functions, as long as the variational energy obtained is 
close -but not terribly close as in the straightforward variational approach- to the exact 
ground state energy.

The paper is based therefore on the recent numerical advances
 for solving approximately model Hamiltonians on a lattice: the fixed node\cite{ceperley}, and the 
''generalized Lanczos technique''\cite{sorella}, 
that allows to improve systematically the variational 
energy provided by the effective Hamiltonian approach, by 
combining in an efficient way the 
power of the Lanczos variational technique with the ''effective Hamiltonian approach''.
Trough all the paper and pictures we will use ''FN'' to indicate the ''fixed node approach'', 
, whereas  ''SR ''will indicate the . ''stochastic reconfiguration method'' 
used to apply  the ''generalized Lanczos'' scheme.
In the first part we describe the Lanczos technique, then we derive the effective Hamiltonian 
approach in a slightly more general way than the standard ''fixed node'' method. Finally we show that 
the mentioned ''generalized Lanczos method'' represents  a very 
efficient implementation of both the previous  techniques-
 Lanczos and fixed node-  on a lattice. We also  
 point out some slight but important  improvements  and simplifications to the 
most recent formulation of the ''generalized Lanczos scheme''\cite{sorella}.
In the last section before the conclusion we show some example on the t-J model, where the ''effective Hamiltonian approach''  
is clearly  useful, as the pairing correlation functions appear 
to be rather independent 
from the initial variational guess, even for large system size 
 $L\simeq 50$ and small $J/t$.

\section{The Lanczos technique}

\begin{figure}[t]
\begin{center}
\epsfxsize=13cm
\epsfbox{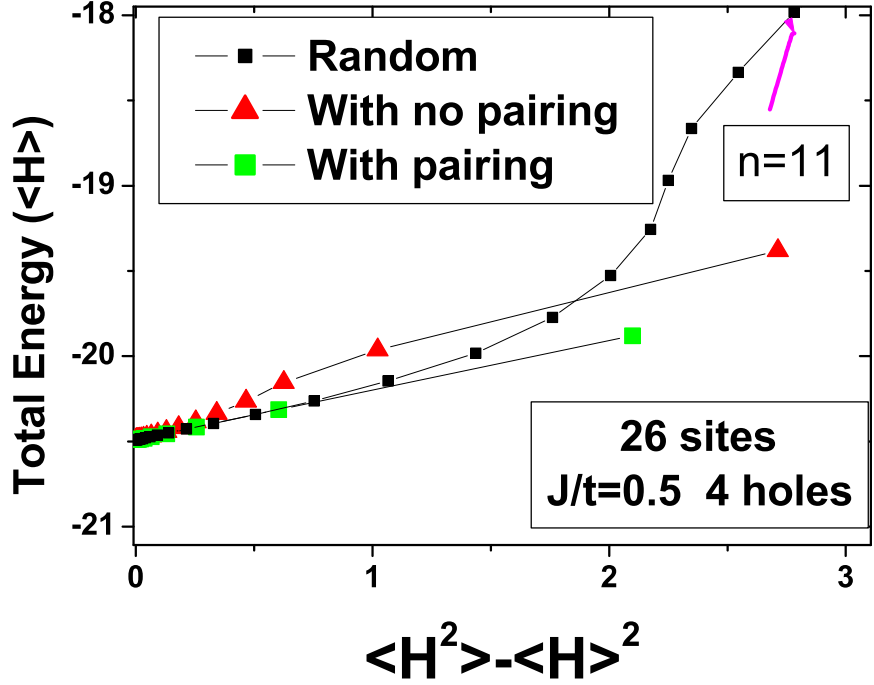}
\caption{\label{lanczos}
Energy $<H>$ vs. variance $<H^2>-<H>^2$  of the Lanczos technique for different initial wavefunction 
$\psi_G$. Here $n$ represents 
the number of iterations. Lower variance is always obtained for larger $n$. The zero 
variance limit is the exact results.
}
\end{center}
\end{figure}

The Lanczos technique represents a  remarkable improvement of the 
power method used to filter out systematically the ground state 
component  of a given initial wavefunction $\psi_G$ by an iterative technique.
The power method is based on the following equation:
\begin{equation} \label{power}
\psi_0 \rangle  \simeq (\Lambda I  -H)^p |\psi_G \rangle 
\end{equation}
where $\Lambda$ is a  suitable large shift to ensure convergence to the ground state for large 
$p$, $I$ is the identity matrix and $|\psi_0 \rangle $ the ground state of $H$.
At a given iteration $p$, after applying just $p$ powers of the Hamiltonian, 
a much better wavefunction $\psi_p$ can be obtained by combining, with proper 
coefficients $\alpha_k$,  the states obtained with the power method 
in the previous iterations:
\begin{equation} \label{psig}
|\psi_{p}\rangle =\Big(1+\sum\limits_{k=1}^p  \alpha_k H^k \Big) 
|\psi_G\rangle
\end{equation}
with parameters $\{ \alpha_k\}$ for $k=1,\cdots, p$ minimizing the energy 
expectation value 
$\langle\psi_p|\hat{H}|\psi_p\rangle / \langle\psi_p|\psi_p\rangle$. 
For any $p$ it is simple to show that the wavefunction (\ref{psig}) 
corresponds exactly to apply 
$p$ Lanczos step iterations to the initial wavefunction $|\psi_G\rangle$.
The $H-$polynomial of degree $p$ which is applied to the initial state $\psi_G$, 
can be generally factorized in terms of its roots $z_i$:
\begin{equation} \label{decompose}
\Big(1+\sum\limits_{k=1}^p  \alpha_k \hat{H}^k \Big)= \prod\limits_{i=1}^p (1-H/z_i) 
\end{equation}
This decomposition will be particular important for applying statistically the  
Lanczos technique with the Stochastic Reconfiguration (see later).
As it is clear from Fig.~(\ref{lanczos}), the Lanczos method converges very quickly 
to the ground state wavefunction especially when a particularly good ''guess'' is 
used for $\psi_G$.  

Whenever the ground state wavefunction is approached 
$|\langle\psi_0|\psi_p\rangle|^2 /\langle \psi_p|\psi_p \rangle^2  = 1 -\epsilon_p $, with 
 $\epsilon_p \to 0$ 
for larger $p$, with 
the energy approaching the exact value with corrections $\simeq \epsilon_p$.
On the other hand, 
the variance $\sigma_p^2$ of the Hamiltonian on the approximate state $\psi_p$ 
$$\sigma^2_p=\langle \psi_p |H^2|\psi_p\rangle -\langle \psi_p |H|\psi_p \rangle^2 = O(\epsilon_p) $$
is going to zero in the limit 
when $\psi_p$ is the exact eigenstate $\psi_0$ 
with the same corrections proportional to $\epsilon_p$.:
It is clear therefore that a very stable evaluation of the energy can be done by using few Lanczos 
steps values of the energy and the corresponding variance. Then, by 
performing simple extrapolation (linear or even 
polynomial), the exact ground state result is  easily estimated  provided the energy-variance 
values are close to the linear regime (see Fig.\ref{lanczos}).
The same scheme can be applied even for correlation functions\cite{sorella}, and represents one of the most 
simple and effective methods to estimate  exact correlation functions with few Lanczos steps 
(i.e. with a minor computational effort)  whenever the variational wavefunction $\psi_G$  is particularly 
good, i.e.  is close to the linear energy vs. variance regime.  Such property of the variational 
wavefunction can be satisfied even for system size $L\simeq 100$\cite{sorella}.

The initial wavefunction to which the Lanczos and the following techniques will be applied 
can be written as follows \cite{gros}:
\begin{equation} \label{gros}
|\psi_G \rangle=|\psi_{p=0} \rangle = \hat{P}_0 \, \hat{P}_N \hat{J} |D\rangle.
\end{equation}
where $|D\rangle$ is a BCS wavefunction, which is an exact eigenstate of the 
following Hamiltonian:
\begin{eqnarray}
\hat{H}_{BCS}&=&\hat{H}_0+ \frac{\Delta_{BCS}}{2} 
(\hat{\Delta}^\dag +\hat{\Delta})  \label{hbcs} \\
\hat{\Delta}^\dag&=& \sum_{\langle i,j \rangle} M_{i,j} \big( 
 \tilde{c}^\dag_{i,\uparrow} \tilde{c}^{\dag}_{j,\downarrow} + 
 \tilde{c}^\dag_{j,\uparrow} \tilde{c}^{\dag}_{i,\downarrow} \label{ddag} \big) 
\end{eqnarray}
where $\hat{H}_0=\sum\limits_{k,\sigma} \epsilon_k \, 
\tilde{c}^\dag_{k,\sigma} \tilde{c}_{k,\sigma}$ 
is the free electron tight binding nearest-neighbor Hamiltonian, 
$\epsilon_k=-2(\cos k_x + \cos k_y) -\mu$, $\mu$ is the free-electron 
chemical potential and  
$\hat{\Delta}^\dag$ creates all possible singlet bonds 
with  d-wave symmetry being $M_{i,j}$, $M_{i,j}$ not restricted 
to nearest neighbors, but exhaustively parametrized with 
a  reasonable number of variational parameters as described in \cite{sorella}.
$\hat{P}_N$ and $\hat{P}_0$ are the projectors over the subspaces with
a fixed number $N$ of particles and no doubly occupied states.
Finally the Jastrow factor 
$\hat{J}=\exp\left( 1/2 \sum_{i,j} v(i-j) 
\hat{n}_i \hat{n}_{j} \right)$ couples the 
holes via the density operators $\hat{n}_i$ and 
contains other few variational parameters.
We note here that by performing a particle-hole transformation on the 
spin down $\tilde{c}^\dag_{i,\downarrow} \to (-1)^i \tilde{c}_{i,\downarrow}$, 
the ground state of the BCS Hamiltonian is just a Slater-determinant with 
$N=L$ particles \cite{shiba}. This is the reason why this variational wavefunction
can be considered of the  generic Jastrow-Slater form, a standard 
variational wavefunction used in QMC. All the mentioned variational 
parameters are obtained by minimizing the energy expectation value of $H$ over 
$\psi_G$\cite{sorella}.

Using the particle-hole transformation, it is also possible to control exactly
the spurious finite system divergences related to the nodes of the 
d-wave order parameter.

\section{The effective Hamiltonian approach}
In a discrete  Hilbert space defined for instance by configurations $x$ 
of electrons with definite positions and spins we consider {\em any} Hamiltonian 
$H$ with real matrix elements $H_{x^\prime,x}$ and {\em any} real 
wavefunction  $\psi_G(x)$  assumed to be non zero for each configuration $x$.

By means of the wavefunction $\psi_G$-hereafter called the guiding wavefunction- we can define a 
two parameter class of Hamiltonians $H^{\gamma}_{FN}$ depending on $\gamma$ and $r$: 
\begin{equation} 
H^{\gamma}_{FN}= \left\{ \begin{array}{crl}  H_{x,x}+ (1+\gamma) {\cal V}_{sf}(x) + r (1+\gamma) e_L(x)
 &{\rm for }~~   x^\prime =&x  \\
  ~~~~~~ H_{x^\prime,x}  ~~~~~~~~~~~~~~~~~~~~~~{\rm if}~x^\prime \ne x   &{\rm and }~~  \psi_G(x^\prime) H_{x^\prime,x} /\psi_{G}(x) <& 0   \\ 
   ~-\gamma H_{x^\prime,x}   ~~~~~~~~~~~~~~~~~~~~~~{\rm  if}~x^\prime \ne x   & {\rm and }~~  \psi_G(x^\prime) H_{x^\prime,x} /\psi_{G}(x) >&0   
\end{array}  \right.
\label{gammaham}
\end{equation}
where the local energy $e_L(x)$ is defined by:
\begin{equation} \label{local}
e_L(x)=\sum_{x^\prime} \psi_G(x^\prime) H_{x^\prime,x}/\psi_G(x) 
\end{equation}
and the so called sign-flip term ${\cal V}_{sf} (x) $ 
introduced in \cite{ceperley} is given by  considering the sum of  
all the {\em positive}   off-diagonal matrix elements appearing in the local energy. 
The effective  Hamiltonian $H^{\gamma}$  has the {\em same} matrix elements of the 
Hamiltonian $H$ for all off-diagonal matrix elements that do not frustrate the guiding function 
signs, the other ones are taken into account by proper modification of the diagonal term.  

The following properties are  almost an immediate consequence of the above definitions:
\vskip 0.5truecm 
\noindent   i)  for $\gamma=-1$ \hskip 1truecm   $H=H^{\gamma}_{FN},$ . 
\newline\noindent  ii) for $r=-1/(1+\gamma)$  and $\gamma \ne -1$  the ground state of $H^\gamma_{FN}$ is the  
  guiding wavefunction itself with zero ground state energy, namely  
$H^{\gamma}_{FN} |\psi_G \rangle =0$. 
\newline\noindent iii) $H= H_{FN}^{\gamma}-(1+\gamma) { d H_{FN}^{\gamma}  \over d \gamma} $
\newline\noindent  iv) $E_L(x)=\sum_{x^\prime} \psi_G(x^\prime) H^{\gamma}_{FN}/\psi_G(x)= e_L(x) (1 +r (1+\gamma)) $  
where $E_L(x)$ is the local energy of the effective Hamiltonian $H^{\gamma}_{FN}$, whereas 
$e_L(x)=\sum_{x^\prime} \psi_G(x^\prime) H/\psi_G(x).$ the corresponding one for $H$. Moreover:
\newline \noindent  (v) for $\gamma\ge  0$ the ground state $\psi_0^{FN} (x) $ of $H^{\gamma}_{FN}$ may be chosen 
to have  the same signs 
of the guiding wavefunction, namely $ \psi_G (x) \psi_{FN} (x)  \ge 0$ for any configuration $x$. 
This follows by doing a unitary transformation of the basis $| \bar x > = { \rm Sign \left[  \psi_G(x ) \right] 
} |x> $, 
in which the off-diagonal matrix elements of the Hamiltonian $H^{\gamma}_{FN,\bar x^\prime,\bar x} <0 $ 
are non-positive.  
Thus the Perron-Frobenius   theorem  holds  implying 
that  a  ground state wavefunction (in principle there maybe degeneracy) can be chosen to satisfy  
$\psi_0^{FN} (\bar x ) \ge 0 $ in the new basis 
, which finally proves (v) in the original basis.  
The statement (v) suggests that the effective Hamiltonian $H^{\gamma}_{FN}$  represents the lattice counterpart 
 of the fixed node (FN) hamiltonian, a  well known approximation  for continuous models.\cite{ceperleycont}
Furthermore, provided the matrix elements of the hamiltonian  $H$ or $H^{FN}$ satisfy an ergodicity 
property (namely that any two arbitrary configurations $x$ and $x^\prime$ can be always connected by 
a suitable large number $M$ of hamiltonian powers $ \langle x^\prime | H^M |x \rangle  \ne 0$), 
then a more restrictive  property holds: the ground state is unique 
for any $\gamma \ge 0$.
This implies immediately that:
\newline \noindent  (vi) the ground state energy $E(\gamma)$ of the   fixed node hamiltonian 
$H^{\gamma}_{FN}$ is an analytic function of $\gamma$,  due to the finite size gap separating 
the unique ground state from the first excited state.   
We assume in the following that this very general property holds for the given hamiltonian 
 a condition which is not restrictive, also considering that 
 if ergodicity is not satisfied, all previous and the following considerations 
hold in all the 
subspaces  of configurations $x$ ergodically connected by the powers 
of the hamiltonian.
\vskip 1 truecm

By using Green Function Monte Carlo the ground state energy $E(\gamma)$  
can be very efficiently computed 
for  $\gamma > 0$  as all the matrix elements of the importance sampled 
Green function $G^{FN}_{x^\prime,x}= \psi_{G} (x^\prime)
 \left[ \Lambda  \delta_{x^\prime,x} - (H_{FN}^\gamma)_{x^\prime,x} \right]/\psi_{G}( x) $ 
are all positive 
for large enough constant shift $\Lambda$. 
This is obtained by averaging the local energy $ < E_L(x)> $ over the configurations 
$x$ generated statistically by the Green function $G^{FN}$ with a standard 
algorithm.\cite{ceperleynand,runge,calandra}  Notice also that,    by property 
(iv),  the local energy $E_L$ of this fixed node hamiltonian 
is proportional to the local energy $e_L$ of $H$   and therefore 
this computation  satisfy the so called 
zero variance property: both $E_L$  and $e_L$ have zero statistical variance 
 if $\psi_G$ is an exact eigenstate of $H$.
 
For $r=0$  $H^{\gamma}_{FN}$   reduces to the standard fixed node hamiltonian 
defined in \cite{ceperley} ( $\gamma=0$)  and extended to
 $\gamma \ne 0$ in \cite{caprio}.
Thus a rigorous theorem holds  
relating the ground state energy $E(\gamma)$  of the fixed node ground state 
$\psi_{FN}^{\gamma}$ of  $H^{\gamma}_{FN}$, to its  variational expectation value 
 $E^{FN} (\gamma) = \langle \psi_{FN}^{\gamma}  |H | \psi_{FN}^{\gamma}  \rangle $ on the hamiltonian $H$:
\begin{equation} \label{fntheorem}
E^{FN} (\gamma) \le E(\gamma) \le \langle \psi_G | H | \psi_G \rangle 
\end{equation}

Using property (i) we therefore  notice that by increasing the value of $r$ from the 
variational value $r=-1/(1+\gamma)$ up to $r=0$ the ground state of the fixed node  
hamiltonian $H_{FN}^\gamma$ becomes a variational state 
with lower energy expectation value.
This implies immediately that the fixed node effective hamiltonian 
is more appropriate to describe the ground state of $H$.

In the continuous case $r$ cannot be extended to positive values because 
the local energy $e_L$ may 
assume  arbitrary large negative values close to the nodes, and   
 the best variational energy can be actually obtained just for $r=0$ 
(since for $r=0$ the fixed node gives the lowest possible energy compatible 
with the nodes of the guiding function).
In a lattice case such a theorem is missing, and  there 
is no reason to expect that $r=0$ is just the optimal value.

 A simple and efficient scheme  to compute a variational 
upper bound of the energy for any $r$ is described in 
 the  following paragraphs.
Using property (iii)  
\begin{equation} 
 E_{FN} (\gamma) = 
  \langle \psi_{FN}^{\gamma} |H^{\gamma} - (1+\gamma) \frac { d H_{FN}^{\gamma} } { d \gamma} 
 | \psi_{FN}^{\gamma}  \rangle  = E(\gamma) - (1+\gamma) { d E(\gamma) \over d \gamma } 
\end{equation} 
where in the latter equality the Hellmann-Feynmann theorem has been used.
By using that  $H^{\gamma}_{FN}$ depends linearly on $\gamma$, 
 the well known convexity property of $E(\gamma)$ holds\cite{lieb} :
\begin{equation} \label{convex}
 { d^2 E(\gamma) \over  d \gamma^2 } \le 0 
\end{equation}
 Therefore the expectation value $E_{FN} (\gamma)$ of the hamiltonian $H$  on the fixed node state 
 is a monotonically increasing function of $\gamma)$, as clearly 
${ d E_{FN} (\gamma) \over d \gamma} = -(1+\gamma) { d^2 E (\gamma) \over d^2 \gamma}  \ge 0 $.
The best variational estimate is obtained therefore for $\gamma=0$, as in the conventional scheme.

The extension to finite $\gamma$ is however convenient to provide better 
variational estimates of $E_{FN}^{\gamma=0}$, which in fact maybe sizable lower than the standard estimate $E_{FN}(0) \le E(0)$ 
for $r=0$. This extension allows also to make a rigorous upper bound of $E_{FN}^{\gamma}$ also in the case 
$r >0 $, without missing  the zero variance property.
In fact, always  by the convexity property of $E(\gamma)$, 
\begin{equation} 
 - { d E(\gamma) \over d \gamma }|_{\gamma=0}  \le -  { E(\gamma) - E(0) \over \gamma} 
\end{equation}
 we finally get that  at the best variational condition  $\gamma=0$ 
\begin{equation} \label{betterbound} 
 E_{FN} (0)  \le E(0)- (E(\gamma)-E(0) )/\gamma. 
\end{equation}
For $r=0$ the above upper bound    improves also the previously known 
value (\ref{fntheorem}), at least for $\gamma$ small enough 
where the  above inequality becomes a strict equality. 
   
\begin{figure}[t]
\begin{center}
\epsfxsize=13cm
\epsfbox{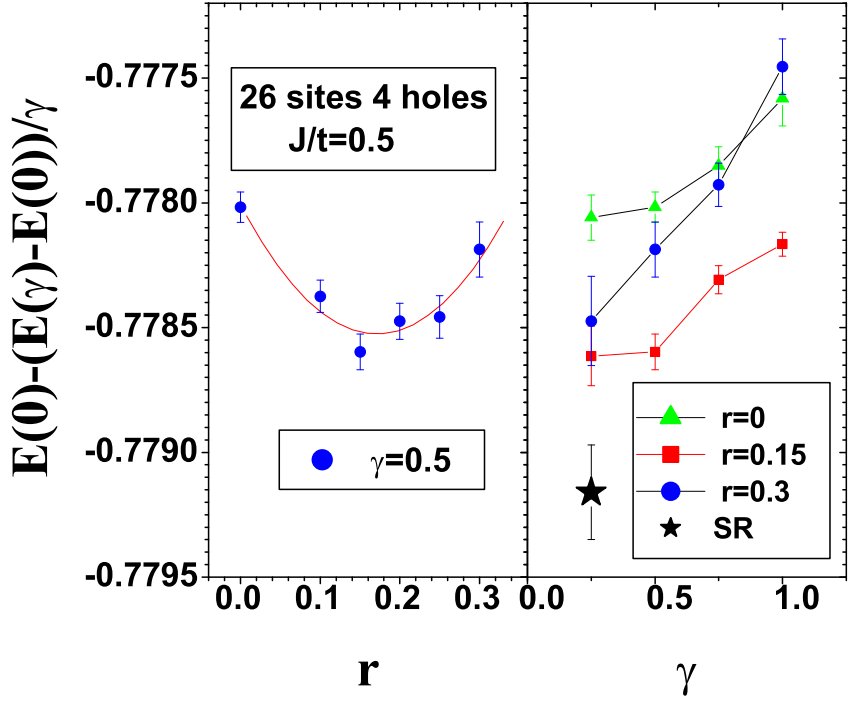}
\caption{\label{fngamma}
Variational energy of the t-J hamiltonian as a function of the parameters $r$ and $\gamma$, for the BCS-guiding function  (\ref{gros}), 
without any Lanczos improvement.
The $\gamma\to 0$ limit  in the right panel corresponds to the expectation value $E^{FN}(\gamma)=
\langle \psi_{FN}^\gamma |H |\psi_{FN}^\gamma \rangle $  for $\gamma=0$
where $\psi_{FN}^\gamma$ is  the ground state of the effective hamiltonian
 $H^{\gamma}_{FN}$. Each point,  
due to inequality (\protect\ref{betterbound}),  represents an upper bound for $E^{FN}(\gamma=0)$ 
and, clearly, for the ground state of $H$. All the estimates reported here 
are  much better than the standard $r=0$ 
lattice fixed node upper bound $E(\gamma=0)$\protect\cite{ceperley} for $E^{FN}(\gamma=0)$:
 $E(\gamma=0)=-0.77580(2)$  much above the upper energy scale. The value (SR) obtained with the ''generalized Lanczos'' described in the following
 sections is also shown for comparison.}
\end{center}
\end{figure}

In practice, since the energy as a function of $\gamma$ is almost linear
   a very good estimate can be obtained using the above inequality 
even  for $\gamma=1$, as shown in  Fig.(\ref{fngamma})  for a test example on the $t-J$ model, 
where it is also  clear that the variational energy can be improved by turning on the parameter $r$.

\section{The generalized Lanczos}
The optimization of the parameter $r$ is rather problematic  within the scheme of the 
previous section especially when few Lanczos steps are applied to the guiding function and the dependence of the 
energy as a function of $r$ cannot be resolved within available statistical errors.
Though the energy maybe rather insensitive to $r$, the behavior of correlation functions, may strongly  
depend on  it, especially when the guiding function shows some instability towards different 
phases not described at the variational level. Within this approach 
the instability of the guiding function is characterized by the existence of a considerable 
number of configurations $x$   with local energy $e_L(x)$ much below the average and with 
correlation properties much different than the average. By increasing  $r$ these configurations 
will have larger and larger weight in the fixed node ground state 
$\psi^{\gamma}_{FN}$ (since they have 
much lower-energy diagonal term)  and will display 
clearly the possible instabilities of the variational wavefunction $\psi_G$.

 The sign-flip term ${\cal V}_{sf} (x) $ 
is divergent whenever the guiding function is exceedingly small (i.e. close to the 
nodes or finite-size lattice pseudo-nodes of $\psi_G$), 
thus requiring an infinite shift $\Lambda$\cite{caprio}, because for the statistical  implementation 
of the power method  the diagonal term 
$\Lambda - (H^\gamma_{FN})_{x,x}
 =\Lambda -H_{x,x}-(1+\gamma) {\cal V}_{sf}(x) - r (1+\gamma) e_L(x)$ 
(see  Eq.\ref{gammaham}) has to be non negative.
For $r=-1/(1+\gamma)$, in the variational case, a better approach,  but similar in spirit,  
 is obtained by sampling\cite{efros}  
the square of the variational wavefunction $\psi_G$ with a different Green function. 
This  following  
importance sampled Green function is used for the statistical implementation of the power method:
\begin{equation}
G^{\gamma}_{x^\prime,x}=
 \left\{ \begin{array}{ccl} {  1 \over z_{x^\prime} } (\Lambda - H_{x,x})  &{\rm for } &  x^\prime =x  \\
- {1 \over z_{x^\prime} }  \psi_G(x^\prime) (H^{\gamma}_{FN})_{x^\prime,x}/  \psi_G(x)     &  {\rm for} &  x^\prime \ne x    
\end{array}  \right.
\label{man}
\end{equation}
where  $z_x$ is a  normalization factor  obtained by setting
 $\sum_{x^\prime}  z_{x^\prime} G^\gamma_{x^\prime,x}=z_x$, namely:.
\begin{equation} \label{defzx}
z_x=\Lambda - e_L(x)+(1+\gamma) {\cal V}_{sf} (x) 
\end{equation}
In this way it is straightforward to show that:
\begin{equation} \label{property} 
\sum_x G^{\gamma}_{x^\prime,x} |\psi_G(x)|^2=|\psi_G(x)|^2
\end{equation}
Thus the importance-sampled Green function $G^\gamma$ maybe used to generate configurations 
that sample  the variational wavefunction square. 
The advantage of the present  approach
is evident since the diagonal term of the Green function  does not contain the sign-flip term, 
and a finite reasonable $\Lambda$ can be used. For instance in the $t-J$ model $\Lambda$ can be set 
to zero.  Instead  a zero shift  is  
not allowed  for the importance sampled Green 
function of the effective hamiltonian itself: 
\begin{equation}\label{fngreen}
G_{FN}=\psi_G(x^\prime) \left[  \Lambda -(H^\gamma_{FN})_{x^\prime,x} \right] /  \psi_G(x) 
\end{equation} 
which performs the same task for $r=-1/(1+\gamma)$, but with   a less efficient infinite $\Lambda$ 
scheme\cite{caprio}.

In the following, within the spirit of the ''effective hamiltonian approach'',   the variational wavefunction 
is improved by  tuning  a parameter $r$ proportional to the local energy,  in order to modify and improve 
the effective hamiltonian $H^\gamma_{FN}$, whose ground state 
is   just   $\psi_G$ for $r=-1/(1+\gamma)$. This parameter is then changed   in order to be  as close as possible 
to the true hamiltonian for $\gamma\ge 0$, when computations free of sign problem are possible.
Indeed in order to improve $H^\gamma_{FN}$ it is very useful to notice that  
$H^\gamma_{FN}=H$, the exact hamiltonian,   for $\gamma=-1$ and  any non-zero  $r$.
Thus at finite positive $\gamma$ an optimal variational parameter $r$ can be used, that on a lattice, maybe 
significantly different from the fixed node value  $r=0$, since this value represents the optimal 
one only in  a continuous model, when there exists a rigorous proof that $r=0$  provides  the 
minimum possible energy.

In order to determine a feasible scheme for the optimization of $r$ in the lattice case,
we need to implement small modifications of the Green function (\ref{man}). We notice that 
there are two important changes of this  Green function  that are easily implemented:

\subsection{One lanczos step improvement}
 In this case  the Green function (\ref{man}) is modified by:
\begin{equation} \label{1ls}
G^\gamma_{1LS}= r_{x^\prime}  G^\gamma_{x^\prime,x} / r_x
\end{equation}
where $r_x=1 +\alpha e_L(x)$.  After applying  statistically 
the  above Green function, after a large number of iterations 
 the configurations $x$, will 
be distributed according to the weight (not necessarily positive):
$$ \psi_G (x) \psi_1 (x)$$
where 
\begin{equation} \label{1lspsi}
\psi_1=(1+\alpha H) |\psi_G \rangle= \sum_x  r_x \psi_G(x) |x\rangle 
\end{equation}
 is the first  Lanczos step wavefunction  as described in 
Eq.~(\ref{lanczos}).  
Since the Lanczos iteration improves the wavefunction and the factor $r_x$ has not a definite sign 
on each configuration $x$, 
it is clear that the phases of the ground state wavefunction are much better represented  by the signs 
of $r_x \psi_G(x)$  rather than by   the ones corresponding to  $\psi_G(x)$.
The parameter $\alpha=\alpha_1/\alpha_0$ can be  determined by satisfying  the 
 SR conditions\cite{sorella}:
\begin{eqnarray}
\langle \psi_G| H ( \alpha_0  + \alpha_1 e_L ) |\psi_n \rangle &=&   \langle \psi_G   H  (\Lambda   -H) | \psi_n \rangle  \nonumber \\
\langle \psi_G | ( \alpha_0  + \alpha_1 e_L )| \psi_n \rangle &=&   \langle \psi_G   |  (\Lambda -H) | \psi_n \rangle 
\label{srcondition}
\end{eqnarray}
where $\alpha_i, i=0,1$ are computed statistically  at 
 any given iteration $n$ in order to improve the $SR$ state $ r_x  \psi_n(x)$,
 until convergence is reached for large  $n$. In this case $\psi_n(x)$ is independent of $n$ 
and statistically equal to $\psi_G$, whereas $\alpha$ will converge (statistically)  to the exact 
one Lanczos step  value.  Once this value is determined      
 the energy  expectation value over $\psi_1$  can be evaluated by 
statistically averaging the local energy $e_L(x)$ corresponding to $\psi_G$ (and not to  $\psi_1$), 
providing a substantial reduction of computational effort. 
In this case, since the value of $\gamma$ is immaterial  for the statistical  averages, it is more convenient 
to use $\gamma=1$, that minimizes statistical fluctuations.

In general, the use of the SR conditions\cite{sorella} allows to obtain the energy and correlation 
expectation values of the 
$p-$Lanczos step wavefunction $\psi_p$, by using a guiding function $\psi_G$ containing only $p-1$ powers 
of the Hamiltonian, e.g. $|\psi_G \rangle \to |\psi_{p-1}\rangle$.
The use of $|\psi_{p-1}\rangle$ as a guiding function 
 for sampling $\psi_p$ may not be the optimal choice.
In  the following we describe a  guiding function  with better nodes 
than $\psi_{p-1}$ but with 
 the same number $p-1$ of hamiltonian powers, that will be used in the following sections whenever 
the method SR will be applied, 

Using the root decomposition (\ref{decompose})  of the 
 $H-polynomial$ defining the $p-$Lanczos step wavefunction $|\psi_p\rangle$, we can single out any  
real root  $z_k$ and similarly to the first Lanczos step case:
\begin{eqnarray} 
\psi_p(x)=r_x \psi_G(x)   ~~~~{\it with } \nonumber \\ 
r_x &=& 1 -e_L(x) /z_k  \nonumber \\
|\psi_G \rangle  & \to & \prod\limits_{i\ne k} (1 -H/z_i)  |\psi_G \rangle 
\label{decompsig}
\end{eqnarray}
The new local energy $e_L (x)$,  obtained with the new guiding function, will keep  into account 
the phases of the $p-$ Lanczos step wavefunction exactly.
In this way, within this decomposition, it is clear that the best guiding function $\psi_G$ of the 
previous form, is obtained by choosing the real root $z_k$ such that:
\begin{equation} \label{condition}
<1-e_L(x)/z_k > 
\end{equation}
is as far as possible (on average over $\psi_G$) from the zero value. 
This condition (\ref{condition}) will minimize the sign changes of  $\psi_G(x)$ to obtain
 $\psi_p (x)=(1-e_L(x)/z_k) \psi_G(x)$, thus providing 
the best possible phases that we can safely obtain with $p-1$ powers of the hamiltonian applied to the 
bare $\psi_G$.
\subsection{Fixed node improvement}
In this case the Green function is modified similarly:
\begin{equation}\label{greenfn}
G^\prime_{FN} = r_{x^\prime} G^\gamma_{x^\prime,x} / {\it Sgn} ( r_x )
\end{equation} 
It is easily obtained that for $r_x=1 -{1 + r (1+\gamma) \over \Lambda} e_L(x) $ and large shift $\Lambda$, 
the effective hamiltonian $H^\gamma_{FN}$ (\ref{gammaham}) is indeed considered, as for $\Lambda \to \infty$
the matrix elements  of $G_{FN}$ (\ref{fngreen}) 
coincide with the ones defined above for  $ \Lambda G^\prime_{FN}$.
up to $O({1\over \Lambda})$.

In particular for $r=0$, and $\gamma=0$ we recover the standard fixed node\cite{ceperley}. Notice also that, if the hamiltonian is free of sign problem 
${\cal V}_{sf}(x)=0$ and the fixed node is exact. Then the choice $r =0$
 provides 
the exact sampling of the 
ground state of $H$ even for finite $\Lambda$, as the factor $r_x$ is proportional to $z_x$ (\ref{defzx}) and simplifies in (\ref{fngreen},\ref{man}). 
\subsection{Generalized Lanczos}
Using the above Green function (\ref{greenfn}), the parameter $r={-(\Lambda \alpha_1/\alpha_0)-1 \over 1 +\gamma}$, 
a single parameter at any order $p$ of the Lanczos iterations, is optimized 
using the SR conditions(\ref{srcondition}) with  $\psi_n $ now depending explicitly on $n$ and 
differing from the initial guiding function $\psi_G$:
$r_x \psi_n(x)= (G^\prime_{FN})^n \psi_G $.
These conditions  provide, as mentioned before, 
$\alpha_0,\alpha_1$ statistically.\cite{sorella}:
However, in this case, the parameter $r$, determined by the SR condition, may not coincide with 
the lowest possible energy condition. 
A further modification of the Green function\cite{sorella}
\begin{equation} \label{interpola}
G^\prime_{\eta} = r_{x^\prime} G^\gamma_{x^\prime,x} / |r_x|^{1-\eta} {\it Sgn} ( r_x ) 
\end{equation}
 that interpolates between the Lanczos limit (\ref{1ls}) for $\eta=0$ 
(when the SR conditions coincide with the Euler condition of minimum energy) and the 
fixed node limit (\ref{greenfn}) for $\eta=1$  allows to overcome this difficulty, as we get closer but not 
exactly equal to the Lanczos limit, and one can obtain even lower variational energies.\cite{sorella}

For the $t-J$ model  
we avoid to consider here this extra-complication, 
since the SR conditions (\ref{srcondition}) 
have been tested to coincide almost exactly with the Euler 
conditions of minimum 
energy (see Fig.\ref{fngamma}) even for $\eta=1$ at least for $\Lambda=0$.  As shown in the same figure 
the SR may also provide a slightly  lower  energy than the corresponding 
one obtained by the best $r$  effective 
hamiltonian $H^{\gamma}_{FN}$, because for small $\Lambda$ the factor 
$r_x$ in Eq.(\ref{greenfn}) may change sign and 
can correct also the phases of the wavefunction 
and not only the amplitudes.  This is also the reason to work with the minimum 
possible shift $\Lambda$.
In principle it is possible to further improve the variational energy and the nodes of the 
sampled wavefunction, by performing the reconfiguration scheme each $k_p$ steps, with an 
effective Green function:
\begin{equation} \label{greenkp}
G^\prime_{k_p} = r_{x^\prime} (G^\gamma)^{k_p}_{x^\prime,x} /  {\it Sgn} ( r_x ) 
\end{equation}
For  $\gamma=1$, it is possible to work with   $k_p >1 $ and with  reasonable statistical fluctuations 
(that increase obviously with $k_p$). By increasing $k_p$   
the factor $r_x$ provides non trivial changes to the phase 
of the wavefunction with corresponding improvement in energy expectation value.
We have not systematically studied this possible modification  of the method so far.
This extension to $k_p>1$  should be clearly useful for model hamiltonians, such as the 
Hubbard model at strong coupling, when a large shift $\Lambda$ is required for 
the convergence of the method.

 For $\Lambda=0$ or finite, the coefficient $r$ in the factor $r_x$ may have little to do with the coefficient appearing in 
$H^{\gamma}_{FN}$,  but, even at finite $\Lambda$, 
 an effective hamiltonian can be still 
defined\cite{sorella}, which is qualitatively similar to $H^\gamma_{FN}$.
 In the following discussions we will not consider the difference between 
the finite $\Lambda$ effective hamiltonian and the infinite $\Lambda$ 
one (\ref{gammaham}) because it is irrelevant for our purposes. 
 
 At each iteration $p$ of the generalized Lanczos  the special  
 guiding function described in Eq.~(\ref{decompsig}) is used, yielding  optimal 
  phases as close as possible to the $p-$Lanczos step wavefunction. 
As far as the remaining parameter $\gamma$, 
 this is  restricted to be positive 
for statistical reasons (no sign problem). 
  Clearly from property (\ref{convex}),
 the smaller is $\gamma$, the better is the variational energy 
but increased fluctuations occurs for 
computing the SR conditions (\ref{srcondition}). 
On the other hand, the  Green-function shift $\Lambda$ has to be taken as 
small as possible,  compatibly with   $\Lambda- H_{x,x} > 0 $
 for any $x$.  in order to further 
improve the efficiency of the power method. 
Within the SR method by minimizing at best the parameters $\gamma$ 
and $\Lambda$ (or increasing $k_p$) we can further improve 
this technique, in a practical scheme.
The optimization of the parameter $r$, since it affects a change in the effective hamiltonian $H^{\gamma}$
is particularly important for correlation functions. Instead all the other parameters (including $\eta$ 
or $k_p$ for instance) may help to obtain 
slightly lower variational energies, but are in general much less important.
 The variational SR results for the $t-J$ model,  described in the following sections,  are obtained 
with  $\gamma=1/4$ and $\Lambda=0$ and refer to the fixed node Green 
function (\ref{greenfn}), 
whereas the symbol  FN will always refer to the standard fixed-node 
case $\Lambda \to \infty$, $\gamma=r=0$.

\section{Results on the t-J model}
We consider the  pairing correlations in the $t-J$ model for square clusters with periodic 
boundary conditions:
\begin{eqnarray} 
P_{i,j;k,l}&=& \langle \Delta^\dag_{i,j} \Delta_{k,l} \rangle  \nonumber \\
 \Delta^{\dag}_{i,j} &=& c^{\dag}_{i,\uparrow} c^{\dag}_{j,\downarrow} + ~~~~ i\leftrightarrow j 
\label{pairing}
\end{eqnarray}
$\Delta^\dag_{i,j}$ creates a singlet pair in the sites $i,j$.
On each lattice we take the first nearest neighbor pair $i,j$ fixed and move $k,l$ 
parallel  or perpendicular to the  direction $i,j$.
In all cases studied the parallel  correlations are positive and the perpendicular ones 
are negative, consistent with a $d-$wave symmetry of the pairing.
The existence of phase coherence in the thermodynamic limit is  obtained whenever 
$P_{i,j;k,l}$ remains finite for large distance separation between  the pair $i,j$  and $k,l$.
A systematic study has been reported in\cite{dagotto}. Here we focus only on few test 
cases to show the power of the method, and the importance to work with an effective 
hamiltonian $H^{\gamma}_{FN}$ with a single variational parameter $r$ as described in the 
previous section.
For all cluster used the distance between pair $i,j$ and pair $k,l$ refers to the minimum one 
between $|R_i-R_k|$, $|R_i-R_l|$, $|R_j-R_k|$ and $|R_j-R_l|$. 
Only for the $6x6$ we use the so called Manhattan distance $|(x,y)| = |x|+|y|$, since the pair 
$(k,l)$ in this case is moved in both perpendicular directions. First the pair $(k,l)$ is translated    
parallel to the $x$-axis up to the maximum distance allowed by PBC, and then (for the $6x6$) 
the pair $(k,l)$ is moved parallel to the $y-$ axis.

\begin{figure}[t]
\begin{center}
\epsfxsize=13cm
\epsfbox{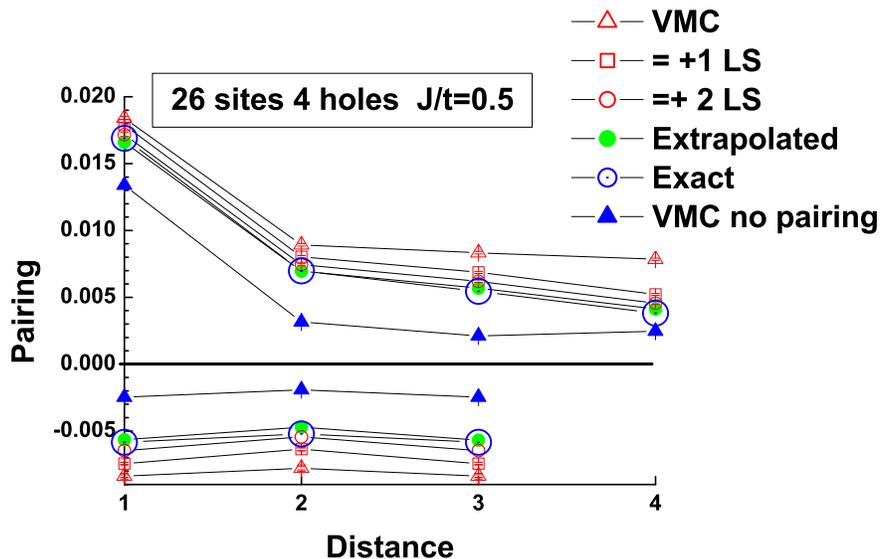}
\caption{\label{pair26}
Pairing correlations in the $26$ lattice for $4$ holes in  the $J/t=0.5$ $t-J$ model 
for the variational Lanczos technique as compared with the exact result obtained 
with exact diagonalization.
The variance extrapolated values are obtained using only the $p=0,1,2$ results 
available with the statistical algorithm also for much larger system size.
}
\end{center}
\end{figure}

First of all, whenever the initial variational wavefunction used is qualitatively 
correct (\ref{gros}), few Lanczos iterations are really enough to obtain exact ground state properties.
This is clearly shown in Fig.(\ref{pair26}) where the exact results coincide within few error 
bars with the variance extrapolated results, that in turn are very close to the 
$p=2$  Lanczos  wavefunction results.
However for larger system when the solution is not known,   
few Lanczos iterations, though systematically improving the energy,   cannot  change 
qualitatively the pairing correlations of the initial wavefunction, and  in general 
the variational approach is not reliable.

In order to show this effect, we have  used two different variational wavefunctions 
on a $6\times6$ 4-holes $J/t=0.5$ cluster, and improved both initializations with the 
methods described in the previous section: the pure variational Lanczos technique, 
the standard fixed node (FN) and the ''generalized Lanczos method'' (SR), 
within the simplified scheme considered before.
For one wavefunction initialization, 
the BCS variational parameters are optimized by minimizing 
the energy, for the other one we have reduced to a very small 
value $\simeq 10^{-4}$  the corresponding variational 
parameter $\Delta_{BCS}$ in (\ref{hbcs}),  just in order to remove
 the degeneracy of the free-electron 
determinant in the $6x6$. This choice yields a variational wavefunction with definite quantum numbers and  with small pairing correlations.

\begin{figure}[t]
\begin{center}
\epsfxsize=15cm
\epsfbox{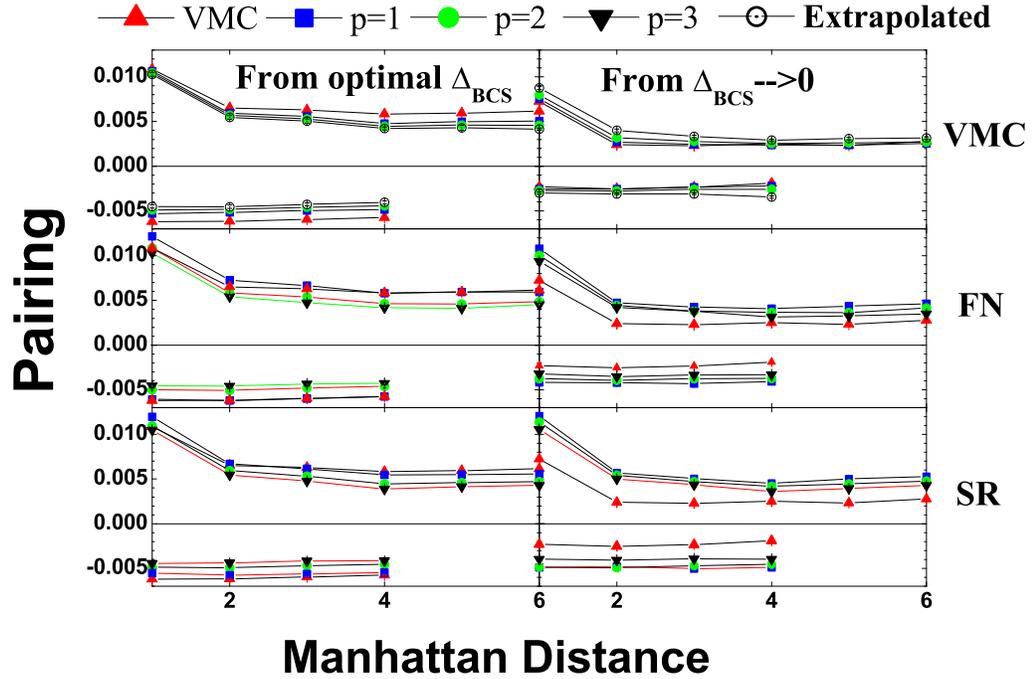}
\caption{\label{pair36}
Pairing correlations in the $6x6$ lattice for $4$ holes in  the $J/t=0.5$ $t-J$ model.
Left panels and right panels refer to two different initial guiding functions with or with 
vanishing  small d-wave order parameter respectively. The latter is used 
in order to remove the degeneracy of the free electron Slater-determinant.
The panels  at different raws refer to different methods, as a function $p$ 
of the hamiltonian powers  used to evaluate the local energy $e_L$, required by  
all the methods: the larger is $p$, the more ( $L^p ~ {\rm for } ~p\ge 2$)
 computationally demanding  is the calculation.
The VMC values (red trangles) are plotted in all panels for comparison.
}
\end{center}
\end{figure}

\begin{figure}[t]
\begin{center}
\epsfxsize=12cm
\epsfbox{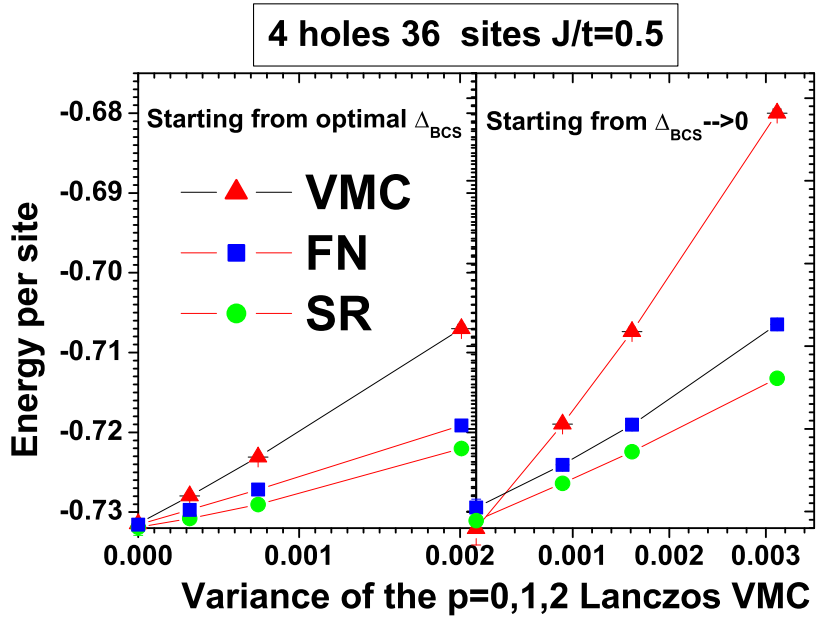}
\caption{\label{ener36} Variational energies obtained with various methods  
as a function of the variance per site $\sigma^2/L^2$  of the $p-$Lanczos step wavefunction (VMC), 
which is improved either  with standard fixed node (FN) or the generalized Lanczos  (SR), with the simplified and efficient scheme described 
in the previous section.
The values at zero variance are extrapolations with a quadratic least 
square fit.
}
\end{center}
\end{figure}

We see in Fig.(\ref{pair36}), top panels, that the Lanczos technique 
is very much dependent on the two different initial choices, even though the energy 
is in both cases very much improved by few  Lanczos iterations.
As shown in Fig.(\ref{ener36}), the variance extrapolated results of the energy 
are consistent for both initial wavefunctions.
On the other hand the pairing correlations remain inconsistent for about a factor two at 
large distance.

In this example we clearly see the limitation of the straightforward variational technique:
within a very similar energy (e.g. the extrapolated ones) the pairing correlations 
maybe  even qualitatively  different. 

A completely different behavior is obtained as soon as the  FN is applied 
(middle panels in Fig.~\ref{pair36}).
The energy improvement within  this technique is apparently marginal compared to the 
standard Lanczos technique (see Fig.~\ref{ener36}).
 Instead  the behavior of pairing correlations is much better, and already the 
simple fixed node approximation applied to the pairing correlations is rather 
independent of the initial wavefunction.
The only drawback of this technique is that when systematic improvements to 
the variational wavefunction are implemented ( larger $p$ in the figure), 
the convergence properties are not behaving  so accurately, as one could expect 
from the convergence of the energy reported in Fig.(\ref{ener36}). 
In particular, even  at the most accurate level of this fixed-node approximation 
-namely the fixed node over the  two Lanczos step wavefunction- the  two different 
initializations give pairing correlations differing by about $20\%$ at the 
largest distance. This is 
much better than the straightforward Lanczos 
variational technique (this difference was about $70\%$ for the corresponding 
 two Lanczos step wavefunctions) but is not satisfactory enough.
 
The reason of such behavior is easily understood in terms of the effective 
hamiltonian approach. In a lattice case it appears  really important for correlation  
functions to optimize the parameter $r$ appearing in the effective 
hamiltonian (\ref{gammaham}) and not just taking 
the FN ansatz $r=0$. This   optimization scheme  is particularly 
important whenever some correlations that are not included at the variational 
level (or much depressed as in the case studied) are increasing as we 
go down in energy with the help of  the improved $p-1$ ($p>1$) Lanczos step 
 guiding function. 
 In general for larger $p$  the parameter $r$ 
increases, thus the SR scheme provides correlation functions
 substantially different and more accurate than  the FN.
In the bottom panels it is remarkable  that, after applying only $3-$steps 
of the SR technique, both 
initializations {\em  provide the same results within error bars ($\le 3\%$) 
at the largest distance}.  
These results can be considered benchmark accurate calculations of pairing 
correlations in the $6x6$ cluster. These pairing correlations
   clearly indicate a  robust 
$d-$wave  superconducting ground state in the $t-J$ model,  at least for 
this $J/t$ ratio.
In this example we notice that correlation functions, in the effective 
hamiltonian approach, begin to be consistent within $5\%$ 
whenever the variational energy 
is accurate within $\sim 1\%$, that is at least  one order of magnitude better 
than a straightforward variational technique like the Lanczos one. 

\begin{figure}[t]
\begin{center}
\epsfxsize=12cm
\epsfbox{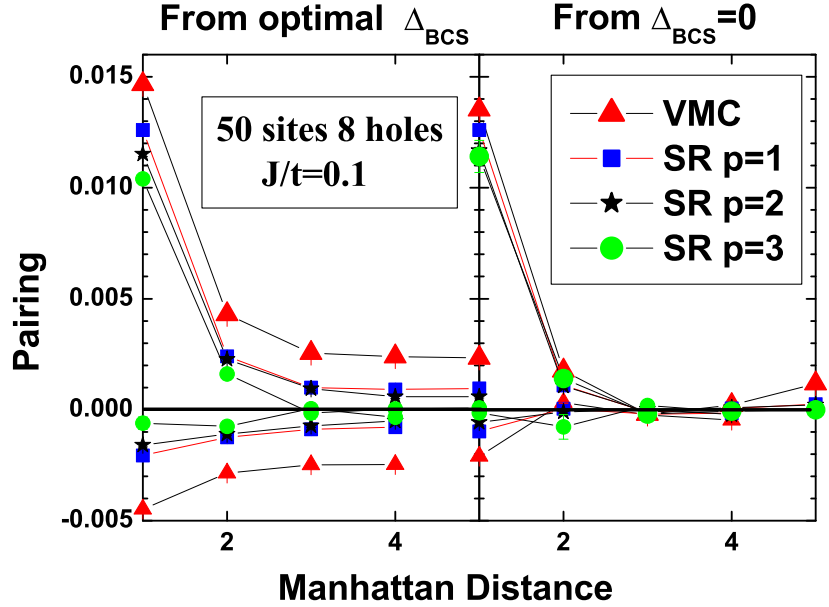}
\caption{\label{pair50}
Pairing correlations in the $50$ site lattice for $8$ holes in  the $J/t=0.1$ $t-J$ model.
Left panels and right panels refers to different initial guiding function with or 
without  d-wave order parameter respectively. 
The pairing correlations for  both calculations are consistently small 
 at the most accurate 
level of approximation (SR $p=3$).
}
\end{center}
\end{figure}

\begin{figure}[t]
\begin{center}
\epsfxsize=12cm
\epsfbox{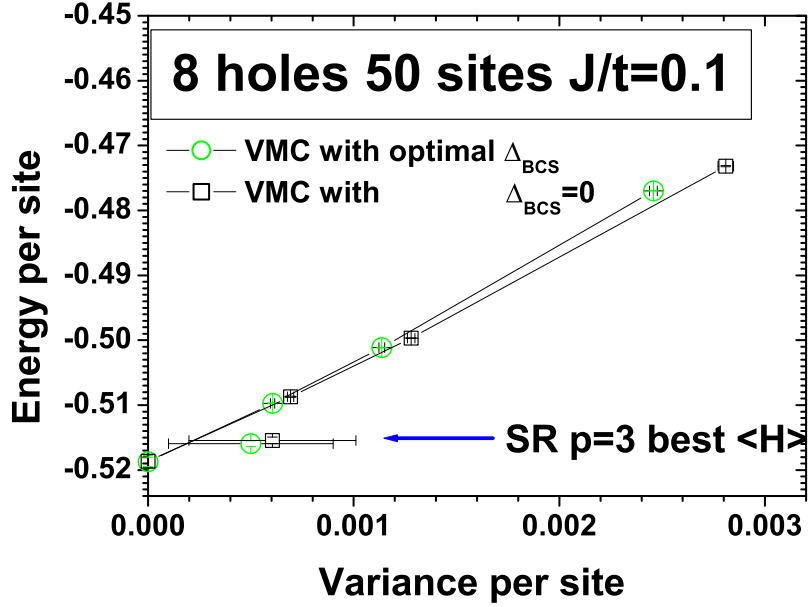}
\caption{\label{ener50}
Variational energy as a function of the variance per site
$\sigma^2/L^2$  for  the $p-$Lanczos step wavefunction (VMC), 
which is improved by the  ''generalized Lanczos  method'' (SR).
The best variational SR $p=3$ energies are indicated by the arrows.
}
\end{center}
\end{figure}

 Of course for larger size,  consistent  correlation functions, i.e. 
independent from the initial wavefunction  with or without $\Delta_{BCS}$, 
 can be obtained for  a larger number $p$ of SR-iterations. 
Here we report a sample case for a 50 site cluster at small $J/t=0.1$.
We see in Fig.(\ref{pair50}) that the sizable pairing correlations present 
in the 
variational wavefunction with $\Delta_{BCS} \ne 0$, represents just an 
artifact of the variational calculation. At the third step, 
of the SR technique, 
when, as shown in Fig.(\ref{ener50}) we reach an accuracy in energy 
below $1\%$ (assuming that the variance extrapolated energies-both consistent- 
are exact), the pairing correlations are again consistent 
within few error bars, 
and clearly vanishingly small at large distance.
   
\section{Conclusions}
We have shown that within a brute force variational technique, such as 
the Lanczos method for few iterations, it is hard to obtain accurate 
values of correlation functions unless the energy accuracy is far 
from the present possibilities, at least in two dimensions. 
 An accuracy of about one part over $10^4$ in the energy would be 
probably possible with at least $~10$ Lanczos steps or $~100000$ 
states in DMRG 
2D calculations  for systems of about $100$ sites with periodic boundary conditions.
This kind of accuracy maybe enough to obtain consistent  correlation functions 
even within these two variational  methods, 
but is far from being possible at present.

We have shown instead that a qualitatively different and very 
promising approach, based on the optimization 
of an effective hamiltonian, rather than adding more variational parameters 
in a brute force variational scheme,   appears to be very useful to control 
correlation functions.  
The idea is based on the ''effective hamiltonian approach'' described in the 
introduction.In this scheme it is  assumed that between 
similar Hamiltonians, the correlation functions of their ground states 
should be also similar. The SR technique, allows to systematically improve 
the effective hamiltonian considered even compared to the 
  lattice fixed node one\cite{ceperley}, 
with an iterative scheme very similar to the Lanczos one, thus the name 
''generalized Lanczos''.

Within this scheme it is clear that there are robust pairing correlations
in the $t-J$ model at sizable but not unphysical value of $J/t$\cite{dagotto}. 
However  there exists a critical value $(J/t)_c \ge 0.1$ 
 below which pairing correlations 
are clearly suppressed. The existence of such a critical $(J/t)_c$  
 is clearly understood  because at $J/t=0$, the 
ferromagnetic instability takes place  even at large doping \cite{ferro}.

\section*{Acknowledgments}
We are indebted to F. Becca, L. Capriotti, A. Parola and  E. Dagotto  
for many useful discussions. 
This work was partially supported by MIUR, 
(COFIN-2001) and INFM (Pais-Malodi).


\begin{thebibliography}{10}

\bibitem{anderson} G. Baskaran, Z. Zou, and P.W. Anderson, 
   \Journal{Solid State Comm.} {63}{973--975} {1987}.
\bibitem{zhang} F.C. Zhang and T.M. Rice, \Journal{Phys. Rev. B} {37}{3759,3763} {1988}.
\bibitem{sorella}
S. Sorella,
{\em Generalized Lanczos algorithm for variational quantum Monte Carlo},
\Journal{Phys.Rev. B}{64}{024512 1--16}{2001}.
\bibitem{white}  S.R. White and D. Scalapino, \Journal{Phys. Rev. Lett.} {80}
  {1272-1275} {1998}; ibidem \Journal{Phys. Rev. B} {60}{753}{1999}.
\bibitem{tklee} C.Y. Shih, Y.C. Chen, and H. Q. Lin \Journal{Phys. Rev. Lett.} {81}{1294-1297}
{1998}.
\bibitem{rvb} L. Capriotti, F. Becca, A. Parola and S. Sorella {\em  Resonating Valence Bond Wavefunctions for Strongly Frustrated Spin Systems} \Journal{Phys.  Rev. Lett.} {87} {098201}{2001}. 
\bibitem{ceperleynand} N. Trivedi and D. M. Ceperley  \Journal{Phys. Rev. B} {41} {4552} {1990}.
\bibitem{ceperley} D. F. B. ten Haaf,  
J. M.J.  van Leeuwen, W. van Saarloos, and D. M. Ceperley,
\Journal{Phys. Rev. B}  {51} {13039} {1995}
\bibitem{ceperleycont} {\em see e.g. } D. M. Ceperley \Journal{J. Stat. Phys.}{63}{1237}{1991}. 
\bibitem{gros} C. Gros, \Journal{Phys. Rev. B} { 38} {931} {1988}.
\bibitem{shiba} H. Yokoyama and H. Shiba, \Journal{J. Phys. Soc. Jpn.} {57} {2482} {1988}.
\bibitem{runge} K. Runge, \Journal{Phys. Rev. B} {45} {12292} {1992}; 
 ibidem \Journal{} {} {7229} {}. 
\bibitem{calandra} M. Calandra and S. Sorella {\em Numerical study of the 
two dimensional Heisenberg model using a Green function Monte Carlo technique 
with a fixed number of walkers}, \Journal{Phys. Rev. B} {57} {11446-11456} {1998}
\bibitem{caprio} S. Sorella and L. Capriotti, \Journal{Phys. Rev. B} {61} {2599} {2000}.
\bibitem{lieb} This follows by writing 
$\gamma= p \gamma_1 + (1-p) \gamma_2$, for any 
$\gamma_1,\gamma_2$ and $0\le p \le 1$, thus finding a variational lower bound 
for $E(\gamma) \ge p E(\gamma_1 ) + (1-p) E(\gamma_2)$ 
because the ground state energy $E(\gamma)$ of $H^{\gamma}$ is certainly 
 bounded by the minimum possible energy that can be obtained by each 
of the two terms in the RHS of the following equation:
$H^{\gamma}= p H^{\gamma_1} + (1-p) H^{\gamma_2}$, 
The above  inequality  represents just the 
convexity property of $E(\gamma)$.  
\bibitem{efros} C.S. Hellberg and E. Manousakis, \Journal{Phys. Rev. B} {61} 
   {11787} {2000}. 
\bibitem{dagotto} S. Sorella, G. Martins, F. Becca, C. Gazza, A. Parola 
and E. Dagotto, {\em Superconductivity in the two dimensional t-J model} 
\Journal{cond-mat/0110460}{}{}{2001}
\bibitem{ferro} F. Becca and S. Sorella {\em Nagaoka Ferromagnetism in the 
Two-Dimensional Infinite-$U$ Hubbard Model}
 \Journal{Phys. Rev. Lett.} {86} {3396-3399} {2001}.
\end{thebibliography}
\end{document}